# Characterization of 3D printed micro-blades for cutting tissue-embedding material

Saisneha Koppaka[a], David Doan[a], Wei Cai[a], Wendy Gu[a], and Sindy K.Y. Tang[a]*

[a]Department of Mechanical Engineering, Stanford University, Stanford, CA 94305, USA
*Correspondence to: sindy@stanford.edu

**Abstract**

Cutting soft materials on the microscale has emerging applications in single-cell studies, tissue microdissection for organoid culture, drug screens, and other analyses. However, the cutting process is complex and remains incompletely understood. Furthermore, precise control over blade geometries, such as the blade tip radius, has been difficult to achieve. In this work, we use the Nanoscribe 3D printer to precisely fabricate micro-blades (i.e., blades <1 mm in length) and blade grid geometries. This fabrication method enables a systematic study of the effect of blade geometry on the indentation cutting of paraffin wax, a common tissue-embedding material. First, we print straight micro-blades with tip radius ranging from ~100 nm to 10 µm. The micro-blades are mounted in a custom nanoindentation setup to measure the cutting energy during indentation cutting of paraffin. Cutting energy, measured as the difference in dissipated energy between the first and second loading cycles, decreases as blade tip radius decreases, until ~357 nm when the cutting energy plateaus despite further decrease in tip radius. Second, we expand our method to blades printed in unconventional configurations, including parallel blade structures and blades arranged in a square grid. Under the conditions tested, the cutting energy scales approximately linearly with the total length of the blades comprising the blade structure. The experimental platform described can be extended to investigate other blade geometries and guide the design of microscale cutting of soft materials.

## 1. Introduction

From food processing to tissue dissection in histology labs, cutting soft materials is of interest in myriad contexts. Cutting tools, including razor blades and knives, are frequently used to induce material failure. Despite its ubiquitous nature, the mechanism of cutting soft materials remains incompletely understood [1–5]. Even in the simple case of indentation cutting of soft materials, where a blade is oriented and pressed perpendicularly into a substrate, many fundamental questions remain, such as how changing the tool geometry changes cut initiation and propagation [6]. In indentation cutting, decreasing the blade tip radius has been shown to be the most significant geometric feature in reducing the energy to puncture the material, [7–9] likely due to a high stress concentration and small material deformation [10]. Razor blades or feather blades are commonly used to study indentation cutting. Commercially, these blades are manufactured to have a tip radius as sharp as ~50 nm. A razor blade is cut out by a pressing machine, and then sharpened by passing through grinding wheels. While it is possible to dull the blade, it is difficult to control the tip radius of commercial razor blades for studies to investigate the effect of blade geometry on cutting.

Previously, we performed indentation cutting of soft biological tissues in the microscale by using a "µDicer" consisting of a grid of silicon micromachined micro-blades (i.e., blades that have a blade length <1 mm). We were interested in micro-blades because of our intent to use them to generate microtissues for organoid culture and other analyses [11]. However, because our micro-blades were formed using a combination of isotropic and anisotropic silicon etching, it was challenging to systematically vary blade geometry by controlling the etch conditions. Recently, high-resolution two-photon polymerization direct laser writing lithography (TPP-DLW) has been applied to fabricate structures down to 100 nm [12–14]. To our knowledge, we were the first group that has applied TPP-DLW to fabricate micro-blades for cutting soft biological materials such as a living cell [15]. However, because cells have heterogeneous mechanical properties, it was difficult to predict the effect of blade geometry on the cutting process in that study.

In this study, to control the blade geometry precisely, we use TPP-DLW to fabricate micro-blades on the Nanoscribe Photonics Professional GT 3D printer. We use paraffin wax as a substrate because it is often used to embed formalin fixed tissue (i.e., FFPE tissue). We focus on paraffin wax rather than FFPE tissue to avoid uncertainties due to tissue heterogeneity. First, we study the effect of blade tip radius (~100 nm to 10 µm) in a single, straight micro-blade on the energy dissipated during indentation cutting of paraffin wax. Second, we test complex blade geometries, specifically blades arranged in a grid, like the configuration in our µDicer. These experiments are possible because we use a 3D printing system to fabricate our blades, and thus we are not limited by the geometric constraints of commercial blades.

## 2. Fabricating 3D printed micro-blades

Figure 1(a) shows the computer-aided (CAD) drawings and scanning electron microscope (SEM) images of a single, straight micro-blade. The blade length, $l$, was fixed at 500 µm, the bevel angle was fixed at 30°, and the height $h$ was 125 µm or 587 µm, much larger than the indentation depth of 20 µm in our experiments. Apart from the micro-blades tested in Figure 3, we only varied the micro-blade tip radius, nominally from 50 nm to 10 µm. We used the Nanoscribe to fabricate blades with tip radius >200 nm using a 25x objective in IP-S resin (Nanoscribe GmbH), and blades with tip radius <200 nm using a 63x objective in IP-Dip resin (Nanoscribe GmbH) on glass or silicon wafer as the substrate [16]. The Young's modulus of these resins (2.1 GPa and 2.91 GPa, respectively) are at least 44x higher than the Young's modulus of paraffin wax (~48 MPa; see Note S1) [17,18].

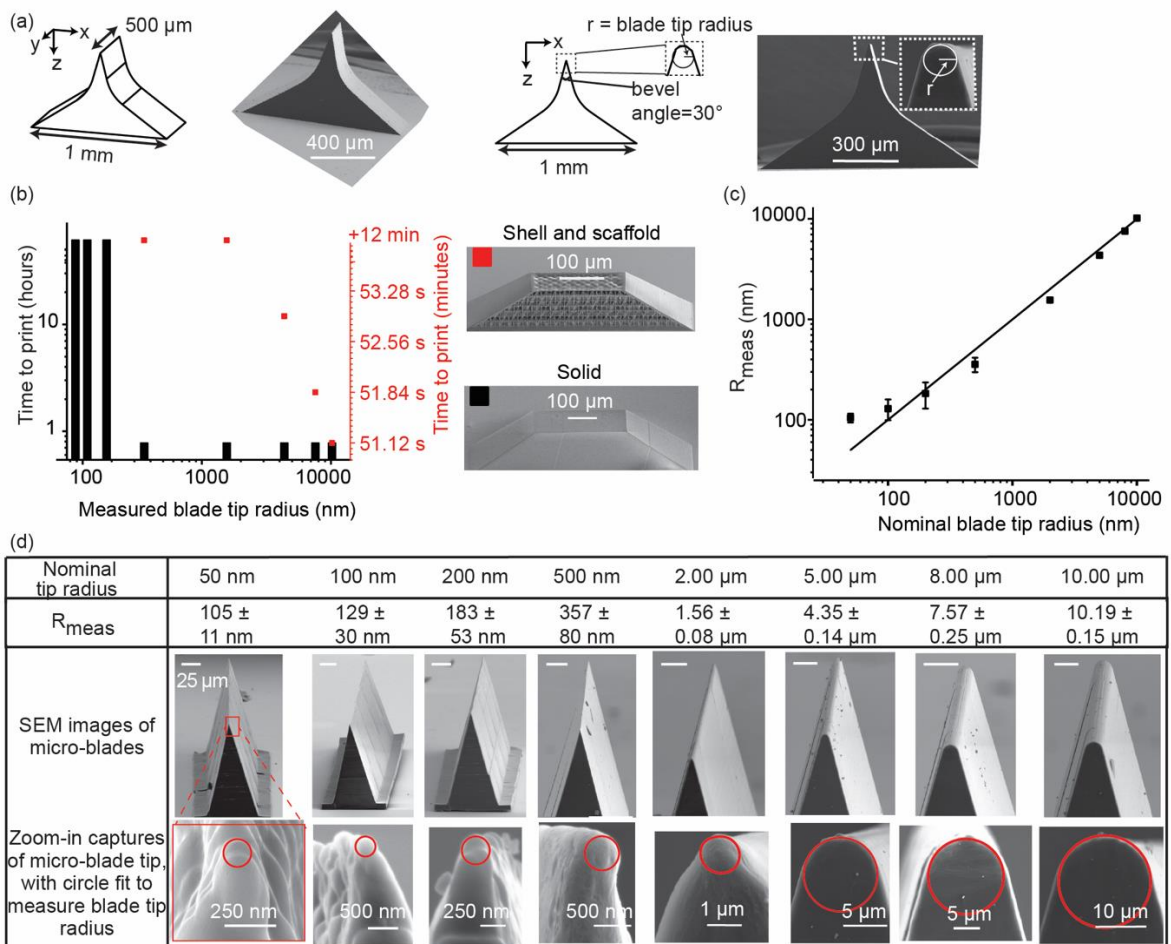

**Figure 1**. (a) Computer-aided drawings (left) and scanning electron microscope (SEM) images (right) of an isometric (top) and a front (bottom) view of the micro-blade. Inset in the SEM image of the front view shows how we measure the blade tip radius. (b) Print times for blades printed in the shell and scaffold mode (black) and solid mode (red) on the Nanoscribe GT system. Insets show SEM images of cross sections of solid and shell and scaffold prints, respectively. (c) Measured blade tip radius $R_{meas}$ vs. nominal blade tip radius (as designed in the CAD). Square markers indicate the average $R_{meas}$. The height of the error bar indicates standard deviation of the measurements for 2-4 printed micro-blades per nominal blade tip radius. (d) A table of nominal and $R_{meas}$ with the standard deviation of the measurements. SEM images and zoom-in capture of each micro-blade. The red circle represents a fit of a circle to the micro-blade tip to measure $R_{meas}$.

Figure 1(b) shows that micro-blades printed in shell and scaffold mode (i.e., a solid outer shell and an inner scaffolding) had shorter print times than those printed in solid mode (i.e., a completely solid structure). For blades with tip radius >200 nm, the print time dropped from >50 minutes using solid mode to <13 minutes using shell and scaffold mode. Thus, we used shell and scaffold mode for these blades. For blades with tip radius <200 nm, the shell and scaffold mode yielded a larger deviation between the nominal and measured tip radius than printing in solid mode. Therefore, we printed these blades in solid mode. However, printing a micro-blade with a height of 587 μm like the one shown in Figure 1(a) would require ~65 hours in solid mode with the 63x objective. To reduce print time, we decreased the blade height to 125 μm for these blades so to reduce the print volume by ~97% and the corresponding print time to 1.5 hours per blade. To increase adhesion between these prints to the substrate, we included a base to the micro-blades (Figure S1). Figures 1(c)-(d) show that the nominal blade tip radius matched well with the printed blade tip radius as measured in SEM (referred to as $R_{meas}$ herein), for $R_{meas} \gtrsim 100$ nm. This length scale is consistent with the voxel size in the Nanoscribe (Figure S2) [16].

## 3. Measuring cutting properties of 3D printed micro-blades

Figure 2(a) shows our setup to measure the cutting properties of the micro-blades. Because of the small length scale of our micro-blades and the soft wax substrate used, we anticipate the range of cutting forces to be smaller than the lower detection limit of a standard mechanical analysis instrument. Therefore, we customized a nanoindenter (Nanomechanics, Inc.) with a lower detection limit <200 nN for load and depth measurements. We mounted our micro-blade perpendicular to a custom nanoindenter tip (see Figure S3). For all experiments, we loaded the blade at a constant displacement rate of 200 nm/s and collected data at 500 Hz. We found insignificant differences in the cutting characteristics among different displacement rates, suggesting the cutting characteristics were not affected by viscoelastic effects of the wax substrate under the conditions tested (Figure S4(a)). The maximum indentation depth was 20 μm into the wax, larger than the micro-blade tip radius (≤10 μm), and much smaller than the thickness of the paraffin wax substrate (~2 mm). We used one large sample of paraffin wax (Electron Microscopy Sciences, 62580-01) in this study to ensure the thickness and the material properties of the substrate were the same in all measurements. We took precautions to ensure our micro-blade was perpendicular to the wax substrate (see details in Note S2).

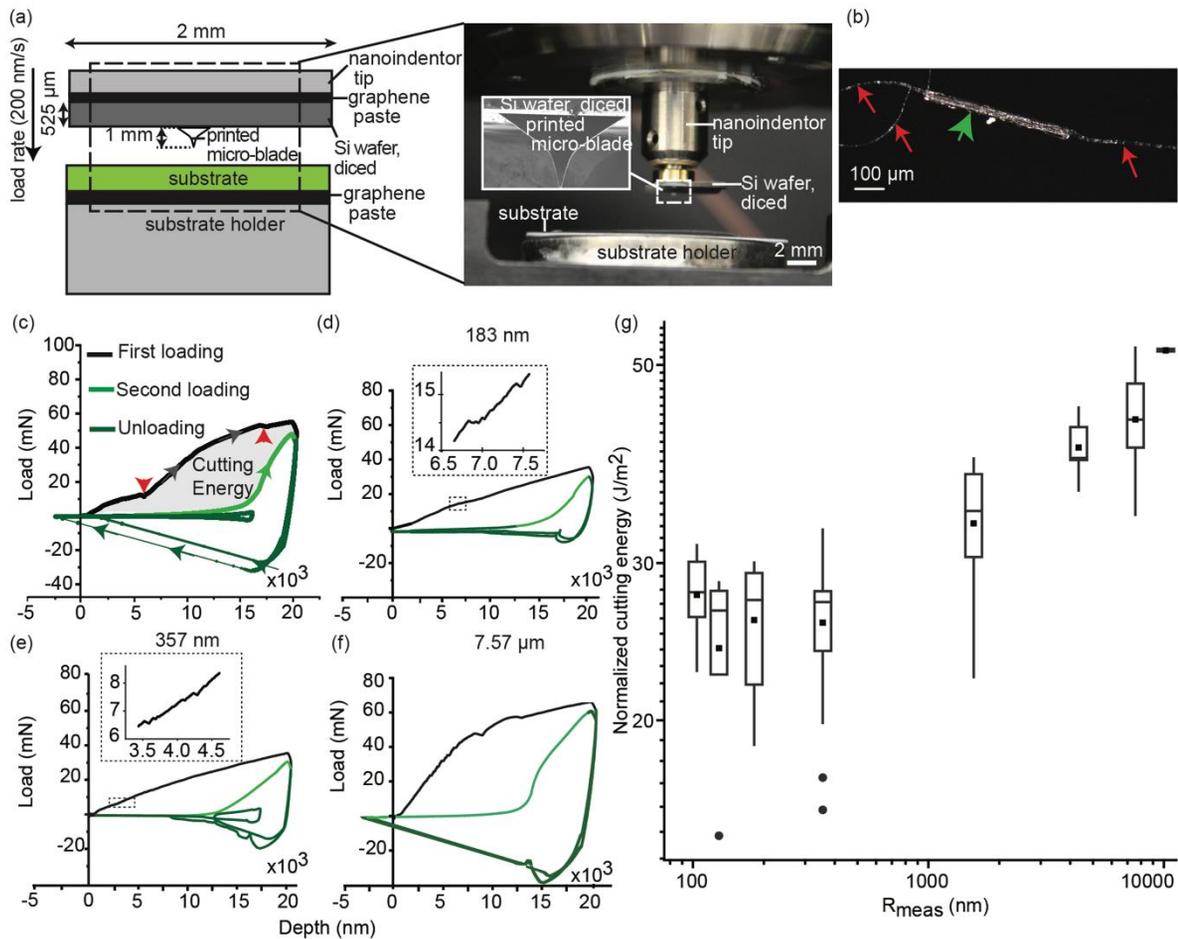

**Figure 2.** (a) Schematic diagram and photograph of our custom nanoindenter setup. (b) Image of wax substrate painted with black Sharpie ink after one loading/unloading cycle using a micro-blade with a tip radius of ~4.35 μm. Green arrow points to the cut made by the micro-blade. Red arrows point to secondary cracks. (c) Example of load-depth data obtained from the nanoindenter using a micro-blade with a tip radius of ~4.35 μm. Black line indicates the first loading, light green line indicates the second loading, and the dark green lines indicate unloading. Arrows indicate the direction of the loading, and red arrowheads indicate saw-tooth events in the first loading curve. (d-f) Representative load-depth data for one run of indentation cutting for four micro-blades with $R_{meas}$ as indicated. One run represents a two-cycle loading/unloading cycles. (g) Box-and-whisker log-log plot of normalized cutting energy (see text for definition) vs. $R_{meas}$. The plot represents the distribution of data across runs of indentation cutting of paraffin wax, using one to five micro-blades at each $R_{meas}$ (see Table S1 for details). Each run was performed at distinct locations on the wax. Square dot represents the mean of the data; whiskers are calculated as first quartile/third quartile ± 1.5*IQR. If data falls within 1.5*IQR, a whisker is drawn up to the data point. If there is an outlier data, a whisker is not drawn to the outliner point; instead, the outlier is represented by a circle.

Figure 2(c) shows an example of one run consisting of two loading and unloading cycles. During the first loading, we observed a saw-tooth pattern along the loading curve, as indicated by the red arrowheads in Figure 2(c). This pattern was not observed in subsequent loadings. Lake and Yeoh reported similar saw-tooth patterns at medium to high loading rates in cutting rubber [5]. They called the behavior "stick-slip" motion and reported that the motion was caused by the substrate material splitting ahead of the blade. Zhang et al. also observed "stick-slip" patterns in the cutting of polydimethylsiloxane (PDMS) [5,19]. In our system, we speculate that stick-slip behavior may be related to fracture in our wax substrate and/or surface friction. However, it is out of the scope of our current study to investigate the origin of the stick-slip behavior.

During the second loading, we observed that load values increased above zero only for displacement values greater than ~12-13 µm. This observation could be explained by the nanoindenter recontacting with the paraffin wax around ~12-13 µm of indentation depth. Unlike the first cycle, we did not observe any stick-slip behavior in the second cycle. Furthermore, the second loading curve followed a characteristic nonlinear J-shaped profile of the load-depth curve, where the slope of the load-depth curve increases with increasing indentation depth. This profile is consistent with the indentation, without cutting, of many soft materials which also exhibit similar J-shaped curves due to the increasing contact area between the indenter and the substrate with increasing indentation depth [8,20,21].

During both the first and second unloading, the load values became negative. We also observed a secondary hysteresis loop during the second unloading cycle. The negative load values likely arose from adhesive forces between the paraffin wax and the micro-blade, while the secondary hysteresis loop likely arose from the nanoindenter actuator attempting to correct for its position [22]. Figure S4(b) shows a load-depth curve with a third loading and unloading cycle. The third cycle had a similar profile to the second cycle, but the profile was shifted slightly to the right, likely arising from thermal drift of the nanoindenter between indentation cycles.

Because the difference between the second and the third cycles was small, we determined the energy dissipation during indentation cutting of paraffin wax between the first and second loading (i.e., using the two-cycle loading method) for our subsequent analysis to compare micro-blades with different tip radii. The two-cycle loading method is a widely used technique that assumes alignment in contact area and contact forces between the two loadings [23,24]. In our system, after the first indentation cycle, the nanoindenter unloaded in the z-direction and reloaded until it recontacted the paraffin wax substrate. Since our nanoindenter did not have any lateral displacement, other than negligible displacement due to thermal drift, we expected alignment in the contact area and contact force between the two loadings. We subtracted the area under the second loading curve from the area first under the first loading curve to determine the energy dissipated (referred to as "cutting energy" herein) due to the creation of new surfaces and plastic deformation of the substrate during the cutting process [7,24,25]. We limited our calculation of cutting energy to positive load values because the negative load values arose from complex interactions between the micro-blade, paraffin wax, and nanoindenter actuator and were not directly related to the cutting process. Nevertheless, including the negative load values increased the cutting energy by <6% and it did not change the trends of subsequent analyses. To verify new surfaces were formed, we used a separate wax substrate with black ink painted on the top surface. We observed that the ink was absent in the cut area after one loading cycle, suggesting new surfaces were formed (Figure 2(b)). We also observed secondary cracks branching from the primary cut surface, consistent with prior observations on indentation cuts [26,27].

## 4. Cutting energy and blade tip radius

Figure 2(d-f) show load-depth plots for blades with $R_{meas}$ of 183 nm, 357 nm, and 7.57 µm, respectively. Figure 2(g) plots the cutting energy normalized by half of the total surface area of the blade in contact with the paraffin wax when the blade was fully indented at a depth of 20 µm, as a function of $R_{meas}$. We observed a decrease in the normalized cutting energy as $R_{meas}$ decreased from 10.2 µm to 357 nm. Thus, for $R_{meas}$ > 357 nm, sharper blades cut more efficiently with a lower energy dissipation than blunter blades, as expected. Since we could not identify features from the load-depth curves indicating cut initiation, the cutting energy here included contributions from substrate deformation before cut, cut initiation, and cut propagation.

Normalized For $R_{meas}$ <357 nm, there was no notable decrease in cutting energy as the tip radius decreased. Kountanya & Endres demonstrated that, when cutting carbon steel with a carbide-tipped tool, the cutting force did not decrease for blades with tip radii smaller than 40 μm [28,29]. In the Y-shaped and frictionless cutting of PDMS with razor blades, Zhang et al. also showed that the strain energy release rate stopped decreasing for blade tip radius ≲ 100 nm [19]. In both cases, it was thought that while sharp blades with tip radius below the threshold produced more intense stresses and strains than blunt blades, they did so over a volume that may be too small to encompass the microstructural features that control fracture [29]. Thus, the plateau in cutting energy may be interpreted as the minimum energy required to activate a threshold damage zone [3,19]. In our study, the plateau in cutting energy may correspond to the fracto-cohesive length, a material-specific scale over which dissipative processes occur in soft solids [30,31]. While further research is needed to determine the fracto-cohesive length for paraffin wax, the fracto-cohesive length was found to be smaller than the elasto-cohesive length $l_e$ in other soft materials [32][3]. We estimated $l_e \sim \Gamma/E \sim 10$ μm for paraffin wax (for fracture energy $\Gamma \sim 469$ J/m$^2$ and elastic modulus ~ 48 MPa), [32-33] about 28 times larger than $R_{meas} = 357$ nm when the plateau occurred.

A plateau in cutting energy could also arise if blades with $R_{meas}$ < 357 nm became blunted while cutting. However, we deemed this possibility unlikely, because: (i) except for the blades with $R_{meas} = 105$ nm, the cutting energy did not increase with run number for a given blade (Figure S5); (ii) the blades had similar $R_{meas}$ before and after the cutting experiments as measured by SEM, and they did not show obvious "blunting" that would explain the plateauing in cutting energy for $R_{meas}$ < 357 nm. The exact origin of the plateauing in cutting energy requires further investigation and will be the subject of a follow-up study.

## 5. Cutting energy of unconventional blade configurations

With 3D printing, we were not constrained to the geometric constraints of commercial blades like a scalpel or a razor blade. This capability allowed us to print and test unconventional geometries, such as the geometries in Figure 3(a). Because micro-blades with $R_{meas} = 357$ nm exhibited a minimum cutting energy in wax and were faster to print than blades with smaller $R_{meas}$, we focused on this tip radius for subsequent analysis (also see Figure S6). Figure 3(b) plots the cutting energy as a function of the total length of the blades. As we increased the total length of the blades, the cutting energy increased approximately linearly ($R^2 = 0.83$) with a slope of ~0.50 nJ/μm. Although we expected the small gap between parallel blades or the sharp corners in blades arranged in a cross, square, or grid configuration might lead to increased material stress and deformation, the cutting energy appeared insensitive to these factors under the conditions tested. This result is useful for the design of complex blade geometries for applications such as a micro-blade grid array.

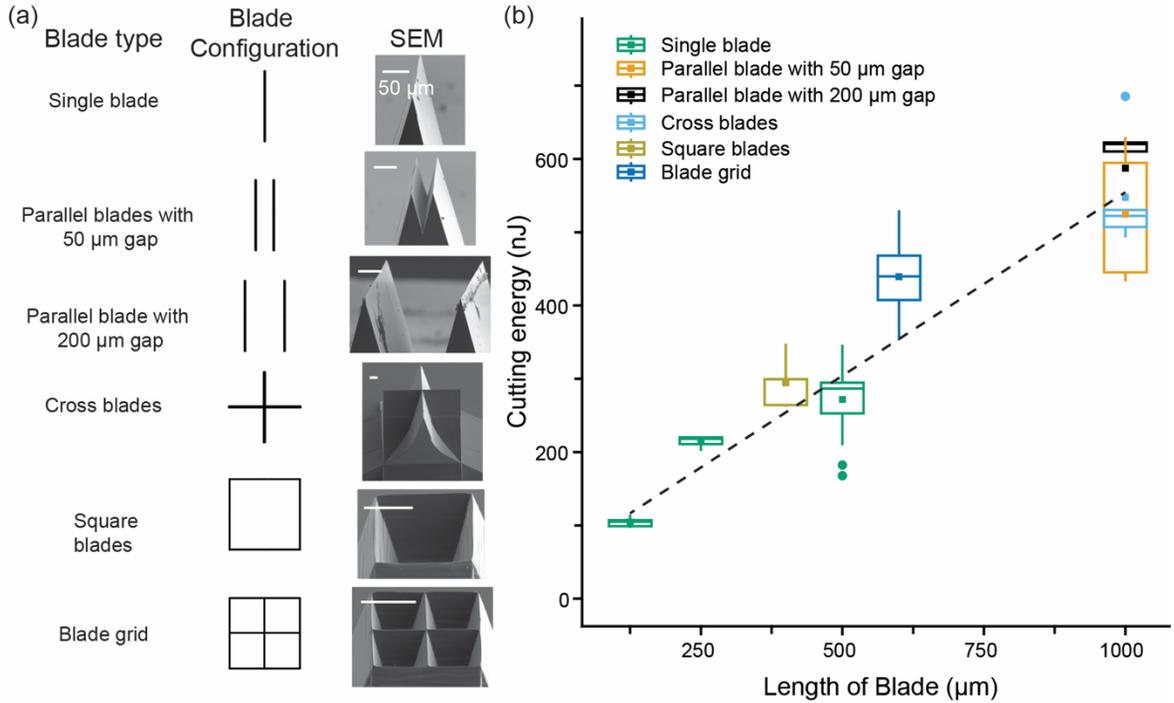

**Figure 3.** (a) Unconventional blade type, configuration, and corresponding SEM images. (b) Cutting energy vs. total length of the blade for all blade types tested. Dashed line shows a linear fit to all data ($R^2 = 0.83$).

## 6. Conclusion

To conclude, we have measured the indentation cutting energy for blades <1 mm in length and with controlled tip radius ranging from ~100 nm to 10 µm for various configurations. Fabricating sharp micro-blades for cutting soft materials is a manufacturing challenge but is made possible with the Nanoscribe 3D printing system. Our results suggest that micro-blades with a tip radius of 357 nm printed in the shell and scaffold mode minimized energy dissipation during indentation cutting of paraffin wax while having a short printing time (<13 minutes). Although our experimental setup does not allow us to differentiate cut initiation from cut propagation under the conditions tested, our results are still useful for applications where these processes are combined. The Nanoscribe allowed us to fabricate unconventional blade grid geometries, and the results have potential to be extended to larger grids with utility in tissue dicing to generate many microtissues at once. Future work includes elucidating the origin of the saw-tooth patterns in the loading curves, the plateau in cutting energy below a tip radius ~357 nm, and the effect of blade grid geometries (e.g., corner effects). Such investigations likely require materials with well-characterized microstructures and/or simultaneous load-depth measurements and imaging of crack formation.

**Acknowledgements**
This work was supported by the Stanford Bio-X Interdisciplinary Initiatives Seed Grants Program (IIP) [R10-14], Stanford Center for Cancer Nanotechnology Excellence for Translational Diagnostics (CCNE-TD) seed grant, Stanford Beckman Center Technology Development Seed Grants, and the National Cancer Institute of the National Institutes of Health (Award Number R21CA261643, R01CA283247). Device fabrication was performed in the Stanford Nanofabrication Facilities (SNF) (NSF Award No. ECCS-1542152). We want to thank Swaroop Kommera at the SNF for his help with the Nanoscribe system, and Brad Ross for valuable discussions on data analysis.

**Declaration of competing interest**
The authors have no conflicts to disclose.

**Author Contributions**

**Saisneha Koppaka**: Conceptualization (equal); Methodology (equal); Investigation (equal); Data curation (equal); Formal analysis (equal); Visualization (lead); Writing – original draft (lead); Writing- reviewing & editing (equal). **David Doan:** Methodology (equal); Investigation (equal); Data curation (equal); Formal analysis (equal); Visualization (supporting); Writing- reviewing & editing (supporting). **Wei Cai:** Investigation (equal); Writing- reviewing & editing (supporting); **Wendy Gu:** Investigation (equal); Writing- reviewing & editing (supporting); **Sindy Tang:** Conceptualization (equal); Methodology (equal); Investigation (equal); Writing – original draft (supporting); Writing- reviewing & editing (equal). Funding acquisition (lead).

**Data Availability**

The data that support the findings of this study are available from the corresponding author upon reasonable request.

# Appendix A. Supplementary Information

**Note S1. Measurement of Young's modulus of paraffin wax sample.**

Using an Instron (Instron 5565), we performed tensile testing on wax samples that were prepared according to Note S2. We measured Young's modulus of paraffin wax, as the slope of the elastic region, to be 47.7 ± 13.7 MPa. This value is similar to values reported in the literature (for example, Wang et al. reported a value of 61.4 MPa in *Adv. Mater*. 18, 1585 (2006)).

**Note S2. Paraffin wax sample preparation and alignment with micro-blade.**

We used a low melting point (56-58 °C) paraffin wax (Electron Microscopy Sciences; Catalog No. 19302-01). This paraffin wax (>99% paraffin with <1% polyisobutylene) was characterized as "ideal for routine histology work," with a more translucent paraffin formulation enabling small specimens to be imaged and sectioned easier. To prepare the paraffin wax substrate, we melted ~5-8 pellets of wax, each pellet ~45 mg, onto a diced silicon wafer piece (~2" x ~2") preheated on a 58 °C hotplate. The wafer with the melted wax pellet was removed and immediately cooled on a metal surface that was pre-cooled to 0 °C using freezer packs.

To ensure that our micro-blade was perpendicular to the paraffin wax substrate, we first used SEM to verify that the micro-blade was printed flat relative to the print substrate. Second, we used the tip of a toothpick to apply a small amount of graphene paste to print substrate. We pressed on the silicon substrate uniformly with tweezers, squeezing out any excess paste, to mount the print substrate to the nanoindenter tip. Finally, to verify that the cut was normal to the substrate, we imaged the indentation on the wax after each cut using a camera attached to the nanoindenter system. We discarded data generated from tilted samples, which showed an indentation having a shorter length or a smaller width than the expected length and width of a normally inserted blade with no tilt.

**Table S1.** Total number of runs (N) and total number of blades used to collect data plotted in Figure 2(g). Each run (N) was performed at distinct locations on the same paraffin wax substrate. Each data point represents one run. In Figure 2(g), the value of $R_{meas}$ plotted is the blade tip radius measured and reported in Figure 1(d).

| | $R_{meas}$ | | | | | | | |
|---|---|---|---|---|---|---|---|---|
| | 105 ± 11 nm | 129 ± 30 nm | 183 ± 53 nm | 357 ± 80 nm | 1.56 ± 0.08 µm | 4.35 ± 0.14 µm | 7.57 ± 0.25 µm | 10.19 ± 0.15 µm |
| Total number of runs (N) | 5 | 5 | 9 | 25 | 10 | 6 | 10 | 3 |
| Total number of blades | 1 | 1 | 2 | 5 | 2 | 2 | 2 | 1 |

**Figure S1**. Schematic diagram of a micro-blade with tip radius <200 nm printed with the 63x objective in solid mode.

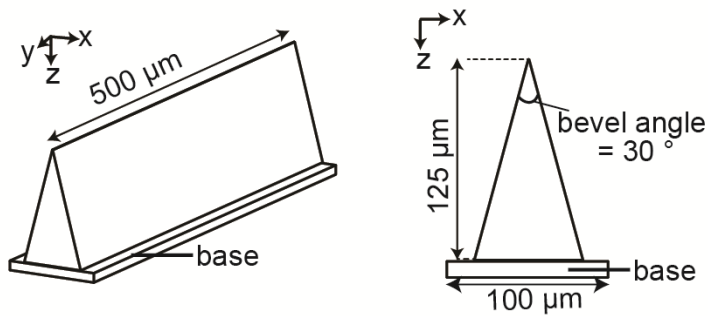

**Figure S2**. Schematic diagram of a print voxel; the minor axis and width of a circle fitted at the tip are labelled. To ensure that only a single voxel comprises the tip of micro-blades with tip radii <200 nm, we checked that only one block rendered the tip in the Nanoscribe DeScribe software. If two or more blocks rendered the tip, we adjusted the block position so that the tip was within a block.

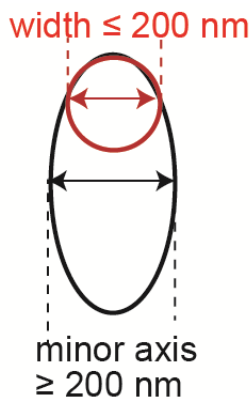

**Figure S3.** Removable portion of the nanoindenter tip to which we graphite pasted our micro-blades.

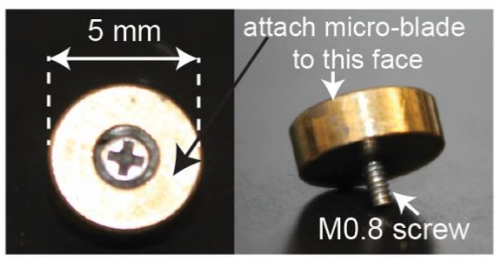

**Figure S4**. (a) Normalized cutting energy for micro-blades with nominal tip radius = 500 nm displaced at 100, 200, 400 nm/s, respectively. We used a permutation test across the data obtained at three displacement rates to compute p = 0.576, suggesting the difference among the displacement rates is not statistically significant. The permutation test is appropriate for testing a global null hypothesis because our sample size is small, and the test makes no assumptions on the distribution from which the data points are drawn. We note several factors could contribute to the run-to-run variability, including variability in the tip radius due to fabrication, micro-blade tilt due to non-uniform application of the graphene paste, and variability in the local mechanical properties in the paraffin wax. (b) Three-cycles of loading and unloading.

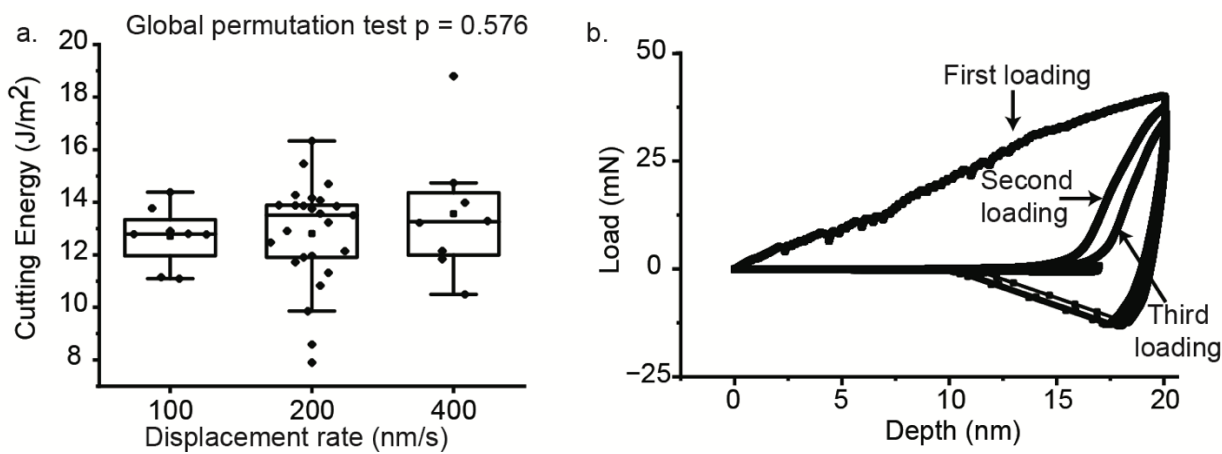

**Figure S5**. Normalized cutting energy vs. run number. For the data shown, only one blade was used for each $R_{meas}$. Except for the blade with $R_{meas}$ = 105 nm, the cutting energy did not increase with run number for a given blade.

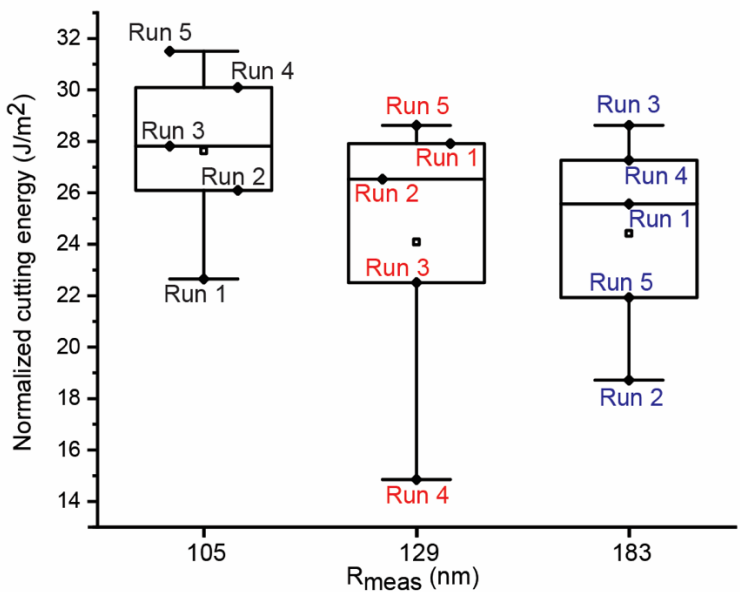

**Figure S6**. Robustness of the micro-blades. While sharp blades may have a low cutting energy, they may be subject to wear and dulling. To quantify the robustness of micro-blades, we indented the blades into a hard material, i.e., aluminum, at 200 nm/s up to a load value of 100 mN, and then indented the blades into our paraffin wax sample and measured the cutting energy of the deformed blade to cut wax. Blades with $R_{meas} <$ 200 nm detached from the substrate or become significantly damaged. Blades with $R_{meas} >$ 200 nm remained largely intact during the aluminum indentation process. These observations indicate that blades with $R_{meas} <$ 200 nm were less robust than blades with $R_{meas} >$ 200 nm, as expected, and we did not perform wax cutting experiments after aluminum indentation for these blades.

Figure S6(a) shows SEM images of the micro-blades after indenting into aluminum and then cutting wax for $R_{meas}$ = (i) 357 nm blade (ii) 1.56 μm (iii) 4.35 μm (iv) 7.57 μm (v) 10.19 μm. The images, labeled (i)-(v), correspond to the box plots in Figure S6(c). The SEM images show a bent tip for the micro-blade with initial $R_{meas}$ = 357 nm but no significant deformation in the other micro-blades.

Figure S6(b) shows representative load-depth plots of micro-blades indented into wax post-aluminum indentation ("post-Al") for a blade with an initial $R_{meas}$ = 357 nm tip radius (top) and a blade with an initial $R_{meas}$ = 4.35 μm (bottom). We observed that, compared with the load-depth plots in Figure 2(c), the plots in Figure S6(b) have a greater area between the first and second load cycles, suggesting greater cutting energy post-deformation. We note the presence of stick-slip events on all first loading curves.

Figure S6(c) shows box-and-whisker plots the difference between the cutting energy after micro-blade was crushed with aluminum averaged for N=1 run ($E_{post\ aluminum}$) and the median cutting energy across all runs ($E_{median}$) for each blade geometry (before crushing into aluminum) as a fraction of $E_{median}$. The difference in cutting energy was the greatest for the micro-blade with initial $R_{meas}$ = 357 nm, suggesting that a micro-blade with this tip radius was less robust than the micro-blades with initial $R_{meas}$ > 357 nm. This result is also consistent with the fact that this blade had a bent tip while other blade tips had no observable deformations (Figure S6(a)).

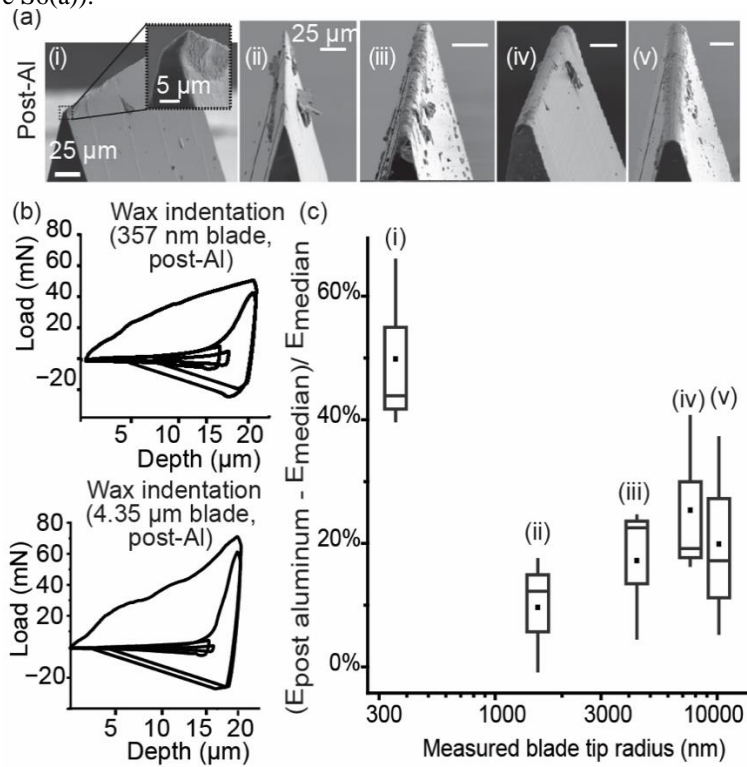